\documentclass[conference,compsoc]{IEEEtran}
\usepackage{times}
\usepackage{fullpage}
\usepackage{comment}
\usepackage[utf8]{inputenc}
\usepackage{graphicx}
\usepackage{color}
\usepackage[colorlinks]{hyperref}
\usepackage{subfigure}
\usepackage{caption}

\usepackage{amsmath}
\usepackage{xspace}
\usepackage{url}
\usepackage{tabularx}

\newcommand{\scenic}{\textsc{Scenic}\xspace}
\newcommand{\verifai}{\textsc{VerifAI}\xspace}

\newcolumntype{M}[1]{>{\centering\arraybackslash}m{#1}}
\title{Addressing the IEEE AV Test Challenge with \scenic{} and \verifai{}}
\author{
Kesav Viswanadha$^\dagger$, Francis Indaheng$^\dagger$, Justin Wong$^\dagger$,\\ Edward Kim$^\dagger$, 
Ellen Kalvan$^\ddag$, Yash Pant$^\dagger$,\\ 
Daniel J. Fremont$^\ddag$, Sanjit A. Seshia$^\dagger$ \\
$\dagger$ University of California, Berkeley \\
$\ddag$ University of California, Santa Cruz}
\date{}

\begin{document}

\maketitle

\begin{abstract}
This paper summarizes our formal approach to testing autonomous vehicles (AVs) in simulation for the IEEE AV Test Challenge. We demonstrate a systematic testing framework leveraging our previous work on formally-driven simulation for intelligent cyber-physical systems. First, to model and generate interactive scenarios involving multiple agents, we used \scenic{}, a probabilistic programming language for specifying scenarios. A \scenic{} program defines an \emph{abstract} scenario as a distribution over configurations of physical objects and their behaviors over time. Sampling from an abstract scenario yields many different \emph{concrete} scenarios which can be run as test cases for the AV. Starting from a \scenic{} program encoding an abstract driving scenario, we can use the \verifai{} toolkit to search within the scenario for failure cases with respect to multiple AV evaluation metrics.
We demonstrate the effectiveness of our testing framework by identifying concrete failure scenarios for an open-source autopilot, Apollo, starting from a variety of realistic traffic scenarios.
\end{abstract}

\section{Introduction}
Simulation-based testing has become an important complement to autonomous vehicle (AV) road testing. It has found a prominent role in government regulations for AVs, for example, one of the National Highway Traffic Safety Administration (NHTSA) missions~\cite{nhtsa} states that AVs should be tested in simulation prior to deployment. Waymo, a leader in the AV industry, has used simulation-based test results to support its claim that its autopilot is safer than human drivers~\cite{waymo}. 

However, there are fundamental challenges that need to be addressed first to meaningfully test AVs in simulation. First, the simulation must effectively capture the complexities of real-world environment, including the behaviors of traffic participants (e.g. pedestrians, human drivers, cyclists, etc), their interactions and physical dynamics, and the the roads and other infrastructure around them. Furthermore, tool support is necessary to (i) specify multiple evaluation metrics with varying priorities, (ii) monitor the performance of the AV according to the specified metrics, and (iii) search for failure scenarios where performance does not meet requirements.

This report summarizes how we formally address these fundamental challenges as participants in the IEEE Autonomous Driving AI Test Challenge. To \textit{model} and \textit{generate} interactive, multi-agent environments, we use the formal scenario specification language \scenic{}~\cite{scenic,scenic-journal}. A \scenic{} program defines an \emph{abstract} scenario as a distribution over \emph{scenes} and behaviors of agents over time; a scene is a snapshot of an environment at any point in time, meaning a configuration of physical objects (e.g. position, heading, speed).
Sampling from an abstract scenario yields many different \emph{concrete} scenarios which can be run as test cases for the AV.
Henceforth we will refer to abstract scenarios simply as ``scenarios''. 

Using \scenic{}, developers can intuitively model abstract scenarios, rather than coding up specific concrete scenarios, by specifying distributions over scenes and behaviors. In conjunction with \scenic{}, we used the \verifai{} toolkit~\cite{verifai} to specify multi-objective evaluation metrics (e.g. do not collide while reaching a destination), monitor AV performance, and search for failure cases by sampling from the distributions in the scenario encoded in \scenic{}. We tested an open-source autopilot, Apollo 6.0~\cite{apollo}\footnote{In the rest of the report, Apollo refers to Apollo 6.0.}, in the LG SVL simulator~\cite{lgsvl} via a variety of test scenarios derived from a NHTSA report~\cite{nhtsa}. Using this architecture, we were able to achieve a high variety of scenarios that highlighted several issues with the Apollo AV system, and we provide some quantitative measures of how well the space of potential scenarios has been covered by our framework.

\begin{figure}
    \centering
    \includegraphics[width=\linewidth]{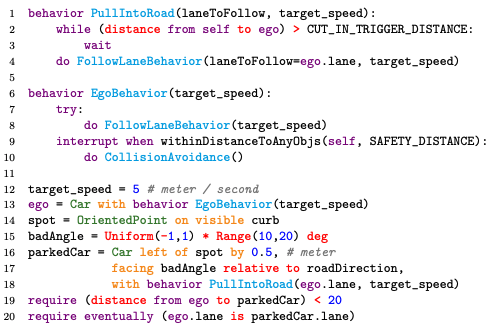}
    \caption{An example \scenic{} program modeling a badly-parked car pulling into the AV's lane}
    \label{fig:scenic_program}
\end{figure}

\subsection{Background}\label{sec:background}

We give here some background on the two main tools previously developed by our research group which we utilize in our approach.

\scenic{}~\cite{scenic,scenic-journal} is a probabilistic programming language whose syntax and semantics are designed specifically to model and generate scenarios. A scenario modelled in this language is a program, and an execution of this \scenic{} program in tandem with a simulator generates concrete scenarios. \scenic{} provides intuitive and interpretable syntax to model the spatial and temporal relations among objects in a scenario.
The probabilistic aspect of the language allows users to specify distributions over scenes and behaviors of objects. The parameters of the scenes and behaviors, from initial positions to controller parameters, form the \emph{semantic feature space} of the scenario. Testing with a \scenic{} program involves sampling concrete scenarios from this semantic feature space. An example of a \scenic{} program, describing a badly-parked car in the ego AV's lane, is shown in Figure~\ref{fig:scenic_program}. Please refer to \cite{scenic-journal} for a detailed description of \scenic{}.

\begin{figure}
    \centering
    \includegraphics[width=\linewidth]{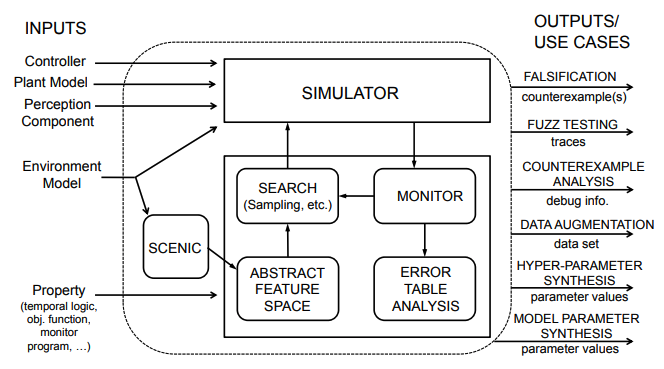}
    \caption{The architecture of the \verifai{} toolkit, from \cite{verifai}.}
    \label{fig:verifai}
\end{figure}

\verifai{}~\cite{verifai} is a software toolkit for the formal design and analysis of systems that include artificial intelligence (AI) and machine learning (ML) components. The architecture of \verifai{} is shown in Fig.~\ref{fig:verifai}. As inputs, it takes the environment model encoded in \scenic{}, system specifications or evaluation metrics, and the system being to be tested. \verifai{} extracts the semantic feature space defined by the \scenic{} model, and searches this space for violations of the specification. Concrete scenarios sampled from the space are executed in an external simulator, while the system's performance is monitored and logged in an error table which stores the results of all simulations. \verifai{} employs a variety of sampling strategies to perform \emph{falsification}, i.e., to search for scenarios that induce the system to violate its specification.
These include passive samplers based on random sampling or Halton sequences~\cite{halton}, as well as active samplers which use the history of the system's performance on past tests to guide search towards counterexamples, such as cross-entropy optimization, simulated annealing, and Bayesian optimization.
Recently, we added a multi-armed bandit sampler which supports multi-objective optimization~\cite{rv21}.

\section{Metrics and Scenarios}

This section presents the safety properties and associated metrics that we use to evaluate an AV's performance, as well as the scenarios these metrics are computed over.

\subsection{Safety Properties and Metrics}\label{eval_metrics}
For all of our generated scenarios, we use the following four safety metrics, based on the evaluation criteria proposed by Wishart et al.~\cite{metrics}. The mathematical formulations for each metric are also given below.

\begin{enumerate}
    \item \textit{Distance}: We require that for the full duration of a simulation, the ego vehicle stays at least some minimum distance away from every other vehicle in the scenario. In our case, the distance between the centers of the AV and any other vehicle must be greater than 5 meters.
    $$\text{always}(\text{distance}(\text{ego}, \text{other}) \geq 5)$$
    \item \textit{Time-to-Collision}: Given the current velocities and positions of the ego and adversary vehicles, the amount of time that it would take the vehicles to be within 5 meters of each other if they maintained their current projected trajectories must always be above 2 seconds. Let $x(t)$ represent the ego vehicle's future position vector as a function of time, using its current position $x_0$ and its current velocity $v_0$, and similarly for another vehicle $x'(t)$ with position $x_0'$ and $v_0'$. Also, let $\|\cdot\|$ be the Euclidean (L2) norm of the input. Then, we have:
    $$x(t) = x_0 + v_0t$$
    $$x'(t) = x_0' + v_0't$$
    \begin{equation}\label{eq:1}
        \| x(t) - x'(t)\| = 5
    \end{equation}
    Let $t_1$ and $t_2$ be the solutions to Equation~\ref{eq:1}, which is a quadratic equation in $t$. We assert that:
    $$\text{always}(\min(t_1, t_2) \geq 2\ \text{or} \max(t_1, t_2) \leq 0)$$
    This corresponds to either a future near-miss/collision being at least 2 seconds away, or being in the past (i.e. the distance between the vehicles is increasing over time).
    \item \textit{Made Progress}: Measures that the ego vehicle moved at least 11 meters from its initial position. This is useful for checking that Apollo is properly communicating with the SVL Simulator, which was not always the case. Let $X$ be an array containing the ego vehicle's position vector at each timestep, for $N$ timesteps. Then, we have:
    $$\| X_1 - X_N \| \geq 11$$
    \item \textit{Lane Violation}: The ego vehicle's average distance from the centerline of its current lane over the entire trajectory must be less than 0.5 meters. Let $\ell_i$ be the center of the current lane at timestep $i$. Using the positions stored in $X$ as defined in the previous metric, this gives us:
    $$\frac{1}{N}\left(\sum_{i = 1}^N{\| X_i - \ell_i \|}\right) \leq 0.5$$
\end{enumerate}

We encode each of these metrics using monitor functions in \verifai{} \cite{verifai}. We then run multi-objective falsification to attempt to find scenarios that violate as many of these metrics at the same time as possible~\cite{rv21}.

\subsection{Evaluation Scenarios}
The test scenarios we selected are scenarios from the NHTSA report~\cite{nhtsa}, encoded as \scenic{} programs. We consider there to be three main components of a scenario, namely: 1) road infrastructure (e.g. 4-way intersection), 2) the type and number of traffic participants (e.g. car, pedestrian, bus), and 3) their behaviors (e.g., lane change). A brief description of the test scenarios studied in this report is shown in Table~\ref{tbl:scenarios}.

\begin{table}[] 
    \centering
    \begin{tabular}{M{0.12\linewidth}|M{0.20\linewidth}|M{0.20\linewidth}|M{0.27\linewidth}}
        \hline Scenario \# & Road Infrastructure & Traffic Participants & Behaviors \\
        \hline 1 & 4-way intersection & 3 sedan cars & AV performs a lane change / low-speed merge \\
        \hline 2 & 1-way road & 2 sedan cars & AV performs vehicle following with a leading car \\
        \hline 3 & 2-way road & 4 sedan cars & AV parallel parks between cars \\
        \hline 4 & 2-way road & 1 sedan car \newline 1 school bus & AV detect and respond to school bus \\
        \hline 5 & 2-way road & 2 sedan cars & AV responds to encroaching oncoming car \\ 
        \hline 6 & 1-way road & 1 sedan car \newline 1 pedestrian & AV detects and responds to pedestrian crossing \\
        \hline 7 & 4-way intersection & 1 sedan car & AV crosses intersection while oncoming car makes unprotected left-turn across path \\ 
        \hline 8 & 4-way intersection & 2 sedan cars & AV makes unprotected left turn at intersection while car from lateral lane cuts across path \\
        \hline 9 & 4-way intersection & 2 sedan cars & AV makes right turn at intersection while car from lateral lane passes \\
        \hline 10 & 4-way intersection & 1 sedan car \newline 1 pedestrian & AV makes unprotected left turn at intersection while pedestrian crosses \\
    \end{tabular}
    \newline
    \caption{Description of the scenarios tested.}
    \label{tbl:scenarios}
\end{table}

\section{AV Simulation Test Scenario Generation}
The purpose of test scenario generation is to search for failure cases while also aiming to cover the test space well. Here, failure is defined by the evaluation metrics used for testing (Sec.~\ref{eval_metrics}), and the test space is comprised of the parameters defined in a \scenic{} program. Any identified failure cases can inform the AV developers of potential susceptibilities of their system. However, biased search towards failures may only exploit a subset of the test space and thereby only provide a partial assessment of the robustness of the system in an abstract scenario. Hence, balancing this trade-off is the key in determining which concrete test scenarios to generate. In this section, we elaborate on our design choices to address this trade-off and highlight a workflow for AV simulation testing. 


\subsection{Scenario Generation Workflow.}
We write a series of \scenic{} programs for the purpose of generating a wide range of concrete scenarios, corresponding to the generic driving situations described in Table~\ref{tbl:scenarios}. We decided which parameters to vary in each scenario based on empirical observation on what seemed to produce and interesting and diverse set of simulations. Some of the parameters that are varied are sampled from continuous ranges, such as speeds and distances from specific points like intersections, whereas others are discrete, such as randomly choosing a lane for the AV's position from all possible lanes in the map file. A given \scenic{} program which encodes an abstract scenario (as defined in the Introduction) is given as input to \verifai{} toolkit, which compiles the program to identify the semantic feature space as defined in Sec.~\ref{sec:background}. Using one of the supported samplers in \verifai{}, a concrete test scenario is sampled to generate a single simulation. Specifically, this sampling consists of \emph{static} and \emph{dynamic} aspects. For each sampling process, an initial scene (e.g. position, heading) is sampled at the beginning and is sent to the simulator for instantiation. During the simulation run-time, distributions over behaviors are sampled dynamically. This dynamic aspect enables generating interactive environment. Specifically, at every simulation timestep, the simulator and the compiled \scenic{} program communicates a round of information in the following way. The simulator sends over the ground truth information of the world to the \scenic{} program, and the program samples an action for each agent in the scenario in accordance with specified distribution over behaviors. These dynamically sampled actions are simulated for a single simulation timestep. This round of communication continues until a termination condition is reached. At this point, the trajectory is input to a \textit{monitor function} in \verifai{}, which computes the values of the safety metrics described in Section 2.2 and determines whether the scenario constitutes a violation or not.

\subsection{Scenario Composition}
An important feature of \scenic{} that we can leverage is the ability to define sub-scenarios that can be composed to construct higher-level scenarios of greater complexity~\cite{scenic-journal}. This facilitates a modular approach to writing simple scenarios that can be reused as the components of a broader scenario. For example, provided a library that includes an intersection scenario, a pedestrian scenario, and a bypassing scenario, \scenic{} supports syntax to form arbitrary compositions of these individual scenarios. Furthermore, we adopt an \textit{opportunistic} approach to composition in which the \scenic{} server invokes the desired challenge behavior if the circumstances of the present simulation allow for it. In the same example, as the ego vehicle drives its path, the \scenic{} server will monitor the environment; if, say, the ego vehicle approaches an intersection, the \scenic{} server will dynamically create the agents specified in the intersection sub-scenario, allow them to enact their behaviors, and subsequently destroy the agents as the ego vehicle proceeds beyond the intersection. This method of composition results in stronger testing guarantees of an AVs performance by providing a sort of integration test that extends beyond the isolated unit test structure of evaluating the AV against scenarios on an individual basis. This yields more opportunities to discover faults in the AV that may otherwise be forgone using only isolated scenario tests.

\subsection{Applied Sampling Strategies}

We navigate the trade-off between searching for failures (i.e. exploitation) and coverage of semantic feature space (i.e. exploration) by using \scenic{}'s ability to write scenarios with parameters that can be searched using any of \verifai{}'s samplers~\cite{scenic-taxinet}.
Specifically, we used Halton and Multi-Armed Bandit (MAB) samplers. The Halton sampler is a passive sampler which guarantees exploration but not exploitation. The MAB sampler is an active sampler that focuses on balancing the exploration/exploitation trade-off~\cite{rv21}.

\begin{itemize}
    \item \textbf{Multi-Armed Bandit (MAB) Sampler.} Viewing the problem as a multi-armed bandit problem,  minimizing long-term regret~\cite{ucb} is adopted as a sampling strategy. This simultaneously rewards coverage early while prioritizing finding edge cases as coverage improves.
    \item \textbf{Halton Sampler.} By dividing the semantic feature space into grids and minimizing discrepancy, the Halton sampler prioritizes coverage~\cite{halton}. Halton sampling covers the entire space evenly in the limit as more samples are collected.
\end{itemize}

\section{AV Simulation Test Results}

\subsection{Scenario classes and test coverage}

\noindent
\textbf{Scenario Diversity:} 
As mentioned in Sec. 2.2, we selected NHTSA scenarios that cover a range of combinations of road infrastructure, traffic participants, and behaviors, and coded these up as abstract scenarios in \scenic{}.  
There are a few factors that contribute to the diversity of our generated test scenarios; first and foremost, every scenario is encoded as a \scenic{} program, which allows for many factors in the generated simulations to be randomized, including the starting positions of the vehicles, their speeds, the weather, and other pertinent parameters of specific scenarios. Furthermore, each non-ego vehicle has a \textit{behavior} associated with it that describes its actions over the course of a simulation. Behaviors themselves contain logic that allows randomization of various aspects of the motion~\cite{scenic-journal}. Because of this, a single \scenic{} program yields a large variety of possible concrete scenarios to generate. \scenic{} and \verifai{}, in turn, sample from this large space of possible scenarios. Our method for approximating a coverage metric of this \scenic{}-defined space of is concrete scenarios discussed below.\par

\noindent \textbf{Coverage Metric:} We implement an estimator for $\epsilon$-coverage~\cite{epsilon_coverage}. The intuition of this metric is the following. A sampled concrete test scenario corresponds to a point in the semantic feature space of a \scenic{} program. After sampling and generating multiple concrete test scenarios, the $\epsilon$-coverage computes the smallest radius, $\epsilon$, such that the union of $\epsilon$-balls centered on each sampled point fully covers the semantic feature space. 

However, this is inefficient to compute exactly, so we approximate this by placing a $\epsilon'$-mesh over the feature space and searching for the finest mesh where every mesh point's nearest neighbor is within $\epsilon'$. This can be implemented by performing nearest neighbor search over sampled semantic vectors with mesh points as queries. We then perform binary search on values of $\epsilon'$ until the search interval is within a given tolerance (set to 0.05 units for our experiments).\par
This metric, qualitatively speaking, tells us how large of an area of unexplored space there is in the feature space. Therefore, the larger the $\epsilon$ value computed by this metric, the less coverage we have of the feature space. One caveat of $\epsilon$-coverage is that it computes the mesh over the entire feature space, which may not necessarily be the same as the \textit{feasible space} of samples --- that is, regions of the feature space which actually lead to valid simulations. This feasible space is usually difficult to calculate exactly \textit{a priori}, and so we make a generalization in this case to use the entire feature space, which is known beforehand. We found that for most of the scenarios, this did not make a huge difference as valid simulations could be found throughout the feature spaces. Another point to note about this metric is that it is computed using only the continuous features sampled by \verifai{}, as computing coverage for sampled variables in \scenic{} is much more difficult given the possible dependencies between variables and their underlying distributions. Therefore, this metric is only an approximation of a subset of all of the features sampled by \verifai{} and \scenic{}.
\subsection{Statistics and distribution report}
For our experiments, we ran several of our abstract \scenic{} scenarios for 30 minutes each, using both the Halton and Multi-Armed Bandit sampler on each scenario. We use a \verifai{} monitor that combines all of the metrics from Section~\ref{eval_metrics}. The results, shown in Table~\ref{tab:all_experiments}, demonstrate that we are able to find several falsifying counterexamples for our various metrics for each scenario.\par
\begin{table}[ht]
    \centering
    \resizebox{.5\textwidth}{!}{
    \begin{tabular}{M{0.10\linewidth}|M{0.10\linewidth}M{0.10\linewidth}M{0.10\linewidth}M{0.10\linewidth}M{0.10\linewidth}M{0.10\linewidth}M{0.10\linewidth}}
        \hline Scenario & Sampler & Total Samples & Progress & Distance & TTC & Lane & $\epsilon$\\
         \hline 1 & Halton & 52 & 0 & 52 & 52 & 14 & --\\
          & MAB & 56 & 2 & 56 & 56 & 54 & --\\
         \hline 2 & Halton & 56 & 6 & 9 & 11 & 0 & 1.245 \\
          & MAB & 53 & 3 & 11 & 11 & 0 & 5.249 \\
         \hline 3 & Halton & 51 & 1 & 51 & 51 & 44 & --\\
          & MAB & 52 & 18 & 52 & 48 & 34 & --\\
         \hline 4 & Halton & 60 & 0 & 55 & 58 & 0 & 0.415 \\
          & MAB & 60 & 0 & 56 & 58 & 0 & 6.079 \\
         \hline 5 & Halton & 76 & 8 & 30 & 30 & 11 & 0.903 \\
          & MAB & 53 & 8 & 47 & 47 & 3 & 49.976 \\
         \hline 6 & Halton & 61 & 2 & 54 & 54 & 2 & 9.497 \\
          & MAB & 60 & 11 & 50 & 50 & 3 & 19.702 \\
         \hline 7 & Halton & 61 & 0 & 0 & 0 & 0 & 0.122 \\
          & MAB & 58 & 0 & 0 & 0 & 0 & 1.099 \\
         \hline 8 & Halton & 57 & 0 & 0 & 0 & 8 & 1.392 \\
          & MAB & 56 & 0 & 0 & 0 & 1 & 3.003 \\
         \hline 9 & Halton & 55 & 0 & 0 & 0 & 1 & 1.294 \\
          & MAB & 53 & 0 & 0 & 0 & 0 & 2.759 \\
         \hline 10 & Halton & 60 & 0 & 5 & 6 & 0 & 8.081 \\
          & MAB & 58 & 0 & 0 & 0 & 0 & 31.128 \\
    \end{tabular}
    }
    \vspace{5pt}
    \caption{The number of samples and property violations found for each scenario, along with the $\epsilon$-coverage metric.}
    \label{tab:all_experiments}
\end{table}
\par

\begin{figure*}
    \centering
    \begin{subfigure}
    \centering
    \includegraphics[width=0.35\textwidth]{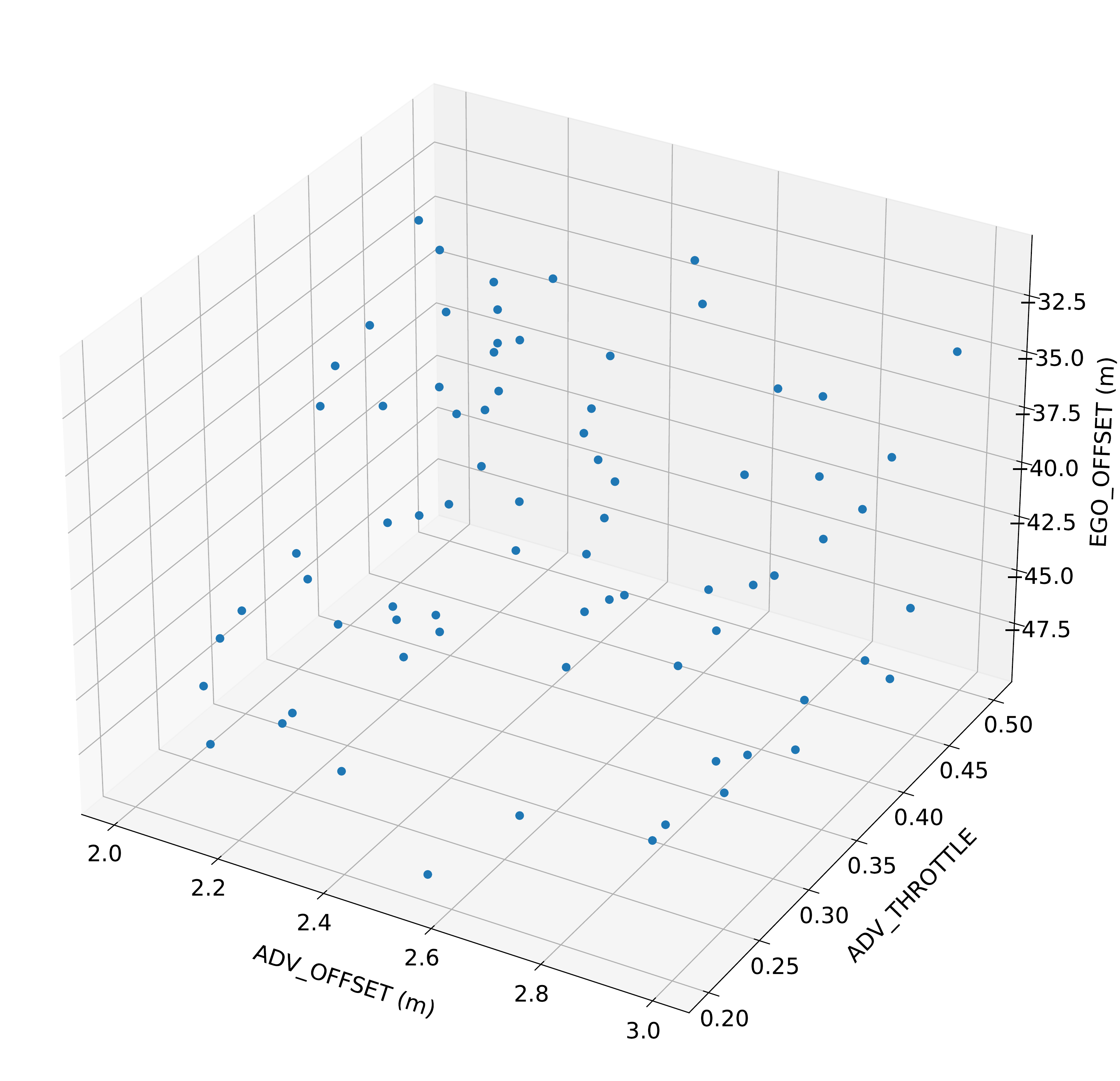}
    \end{subfigure}%
    ~
    \begin{subfigure}
    \centering
    \includegraphics[width=0.35\textwidth]{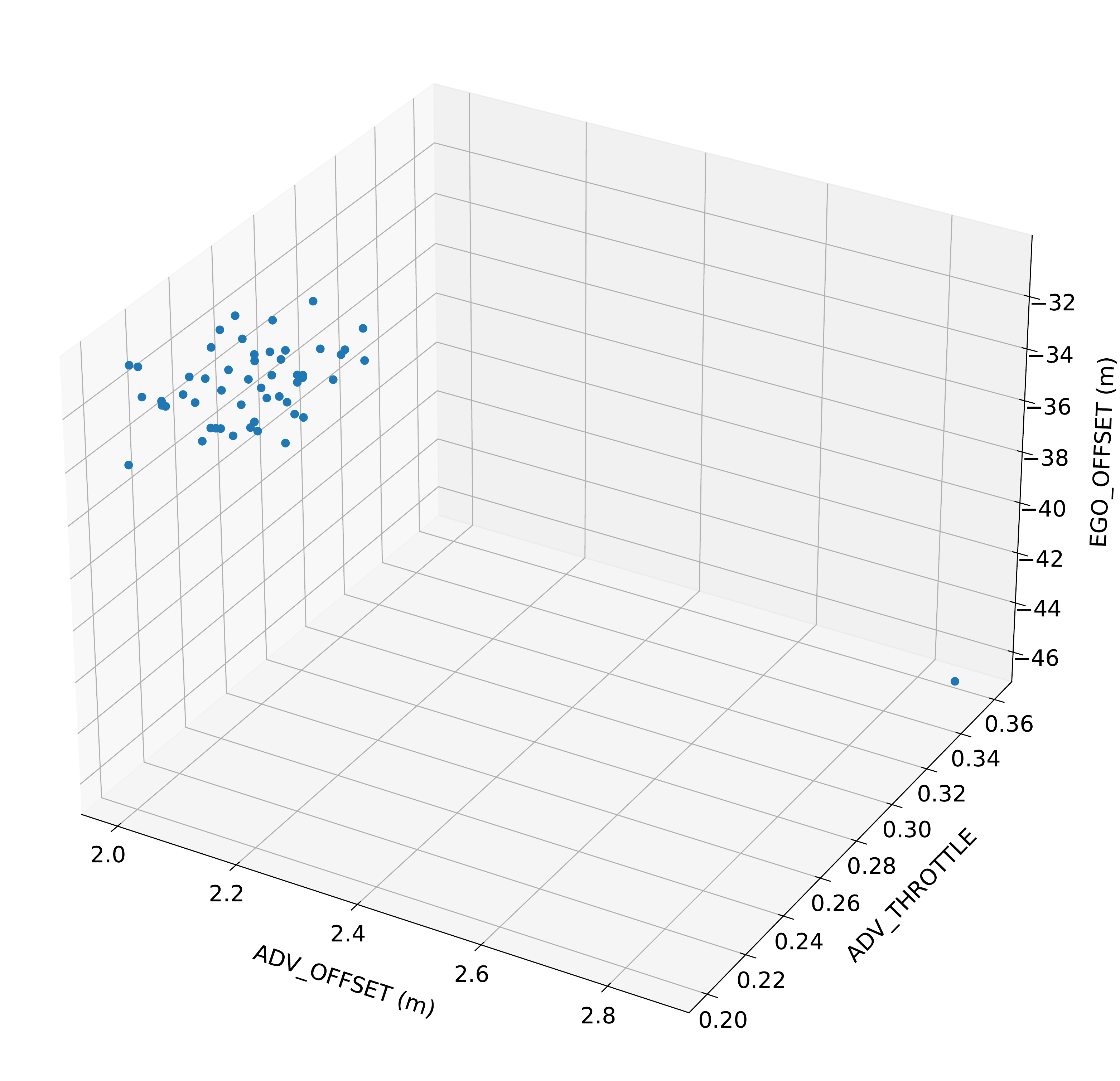}
    \end{subfigure}
    \\
    \begin{subfigure}
    \centering
    \includegraphics[width=0.35\textwidth]{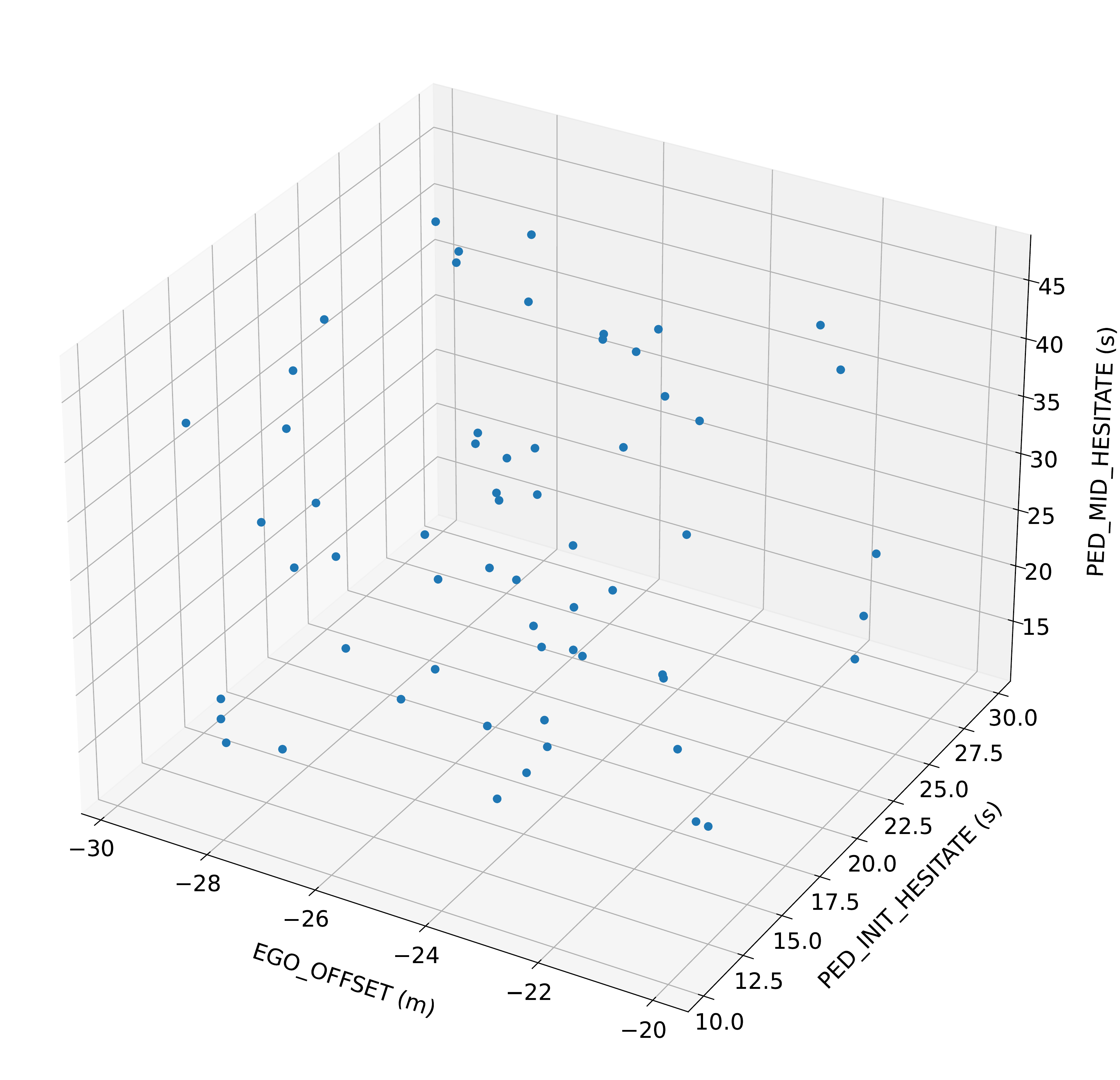}
    \end{subfigure}
    ~
    \begin{subfigure}
    \centering
    \includegraphics[width=0.35\textwidth]{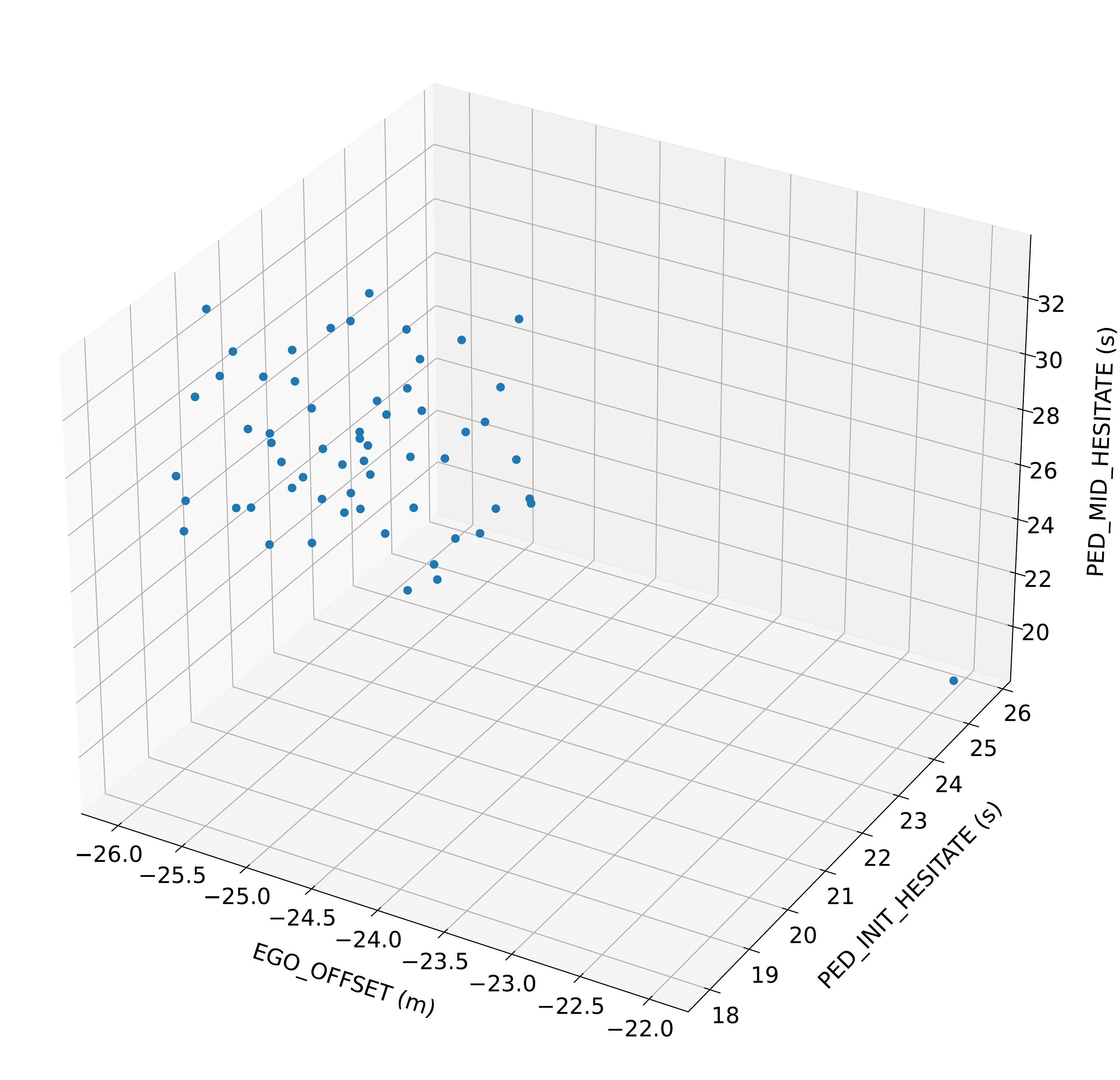}
    \end{subfigure}
    \caption{ The space of points generated by the Halton (left) and MAB samplers (right) for scenario 5 (top plots) and Scenario 6 (bottom plots). In Scenario 5, the axes respectively are the adversary vehicle's distance from the intersection, the adversary vehicle's throttle value, and the AV's distance from the intersection. In Scenario 6, the axes respectively are the AV's distance from the intersection, how long the pedestrian waits before starting to cross, and how long the pedestrian waits in the intersection.}
    \label{fig:point_plot_5}
\end{figure*}

We found that in many of these scenarios, the distribution of safety violations across the feature space was roughly uniform. Because of this, the use of multi-armed bandit sampling did not result in a significantly higher number of violations than using Halton sampling. Moreover, as seen in Fig.~\ref{fig:point_plot_5}, in some of the scenarios multi-armed bandit sampling did not fully explore the search space. We hypothesize that this may have been due to the low number of samples used in our experiments; further investigation is required to understand the MAB sampler's behavior in these cases.

For each scenario, we also present our coverage metric based on the semantic feature space defined in the \scenic{} program. In all scenarios for which the coverage metric was computed, the value of $\epsilon$ is far lower for Halton sampling than it is for MAB sampling, which means Halton gives us better coverage as expected. The plot in Figures~\ref{fig:point_plot_5} reflects the values of these metrics as there is geometrically more empty space in the overall feature space defined by the \scenic{} program for MAB sampling than there is for Halton sampling. The metric was not computed for Scenarios 1 and 3 because there were no continuous-valued features sampled by \verifai{} in those scenarios.

\subsection{Safety violations discovered via AV testing}
Our sampling-based testing approach uncovered multiple safety violations, some of which were a result of the following unsafe behaviors we observed in the simulations:
\begin{enumerate}
    \item In multiple runs of scenarios involving intersections, Apollo gets ``stuck'' at stop signs and doesn't complete turns.
    \item Apollo disregards pedestrians entirely, even though they are visible to the vehicle as shown in the Dreamview UI.
    \item With non-negligible probability, Apollo fails to send a routing request using the \texttt{setup\_apollo} method in the PythonAPI. Because of this, the car does not move much beyond its starting position, hence the \textit{made progress} metric that we included in our experiments.
    \item We also noticed that Apollo sometimes stops far beyond the white line at an intersection, something that would likely present a hazard to other drivers in a real-world scenario.
    \item In the ``vehicle following" scenario (Scenario 2 in Table~\ref{tab:all_experiments}), we were able to find a few dozen samples where Apollo collided with the vehicle in front of it. A few of these examples are also included in our simulation test reports generated by the SVL Simulator.
\end{enumerate}

Depending on the scenario, we found that the number of simulations in which a metric was violated could be quite high. For example, in the pedestrian scenario, almost all of the simulations violated the \textit{time-to-collision} metric, indicating that perhaps the vehicle was approaching the pedestrian much more quickly than what would feel safe to a human passenger.\par
Additional materials, including surce of the Scenic programs, videos of a few simulations, and test reports generated by the LGSVL Simulator, are available at \href{https://drive.google.com/drive/folders/13yQpxdX3mFAUNZnUNUus_rHRFae1h-t5?usp=sharing}{this Google Drive link.}

\section{Conclusion}

We demonstrated the effective use of our formal simulation-based testing framework. Our use of \scenic{} to model abstract scenarios had the following benefits: (i) concisely represent distributions over scenes and behaviors and (ii) embed diversity into our test scenario generation process, which involves sampling concrete test scenarios from the \scenic{} program. To enable intelligent sampling strategies to search for failures, we used the \verifai{} toolkit. This tool supported specifying multi-objective AV evaluation metrics and various passive and active samplers to search for concrete failure scenarios. 
Using our methodology, we were able to identify several undesirable behaviors in the Apollo AV software stack.

Our experiments for the IEEE AV Test Challenge also uncovered some limitations that would be good to address in future work.
For example, running simulations was computationally expensive and limited the number of samples different search/sampling strategies could generate, which seemed to hurt the performance of certain samplers (e.g. the MAB sampler) compared to others from our previous study~\cite{rv21}.



{\small
\bibliographystyle{ieee}
\bibliography{refs}
}

\end{document}